# Gold nanoparticle assemblies: Interplay between thermal effects and optical response

Bruno Palpant,[a] Yannick Guillet,[a] Majid Rashidi-Huyeh,[a,b] and Dominique Prot[a,c]

[a] Université Pierre et Marie Curie – Paris 6, CNRS UMR 7588, INSP, Campus Boucicaut, 140 rue de Lourmel, Paris, F-75015 France
[b] Department of Physics, University of Sistan and Baluchistan, Zahedan, Iran
[c] Université Paris Sorbonne – Paris 4, France
E-mail: bruno.palpant@upmc.fr

## Abstract

The optical response of materials based on gold nanoparticle assemblies depends on many parameters regarding both material morphology and light excitation characteristics. In this paper, the interplay between the optical and thermal responses of such media is particularly investigated under its theoretical aspect. Both conventional and original modeling approaches are presented and applied to concrete cases. We first show how the interaction of light with matrix-embedded gold nanoparticles can result in the generation of thermal excitations through different energy exchange mechanisms. We then describe how thermal processes can affect the optical response of a nanoparticle assembly. Finally, we connect both aspects and point out their involvement in the nonlinear optical response of nanocomposite media. This allows us to tackle two key issues in the field of third-order nonlinear properties of gold nanoparticles: The influence of the generalized thermal lens in the long laser pulse regime and the hot electron contribution to the gold particle intrinsic third-order susceptibility, including its spectral dispersion and intensity-dependence. Additionally, we demonstrate the possible significant influence of the heat carrier ballistic regime and phonon rarefaction in the cooling dynamics of an embedded gold nanoparticle subsequent to ultrafast pulsed laser excitation.

## 1. Introduction

The remarkable optical properties of gold nanoparticles associated with the surface plasmon resonance phenomenon have usually been thought to result in concrete developments in the nanophotonic field. However, one cannot but notice that the main recent breakthroughs have rather been achieved in the domain of thermal applications of these optical properties. Indeed, as optical and thermal responses are in fact closely bound, gold nanoparticles can be considered together as nanometric heat sources and probes for local temperature variations via their optical behavior. The energetic conversion realized by gold nanoparticles which are able to transform at the nanoscale an electromagnetic radiation into heat emitted toward their environment [1] may be indeed relevant in numerous fields. Let us mention some examples. In plasmonic devices [2] local heating may alter the guiding of the electromagnetic wave by gold nanostructures and therefore requires to be well controlled. Gold nanoparticles are also expected to be used in microscopy for labeling biologic cells: Nanoparticle heating by light absorption enables to modify the optical response of their local environment [3]. In the medical area, photothermal cancer therapy based on gold nanoparticles has yet become a very promising technique: assemblies of gold nano-objects (nanorods or dielectric-metal core-shells) absorb light energy transmitted through biologic tissues and transform it into heat which diffuses locally toward local environment. By using an appropriate targeting method for carrying particles close to affected cells, the latter will be destroyed by overheating [4–6]. One may also take advantage of this local heating around particles for inducing local phase or morphology transformation in the surrounding medium. On the one hand, this can enable the measurement of nanoscale heat transfer through the investigation of such phase transformations [1]. On the other hand, this could be used to modify the global medium optical properties [7]. This effect has been supposed to be at the origin of the optical limitation phenomenon in colloidal solutions (induced light scattering by formation of gas bubbles around gold colloids) [8,9]. Metal nanoparticles are also considered as model defects for studying the damage of optical devices induced by powerful lasers [10–12]. The dynamics of the light-heat conversion in a gold nanoparticle and of the thermal release toward its environment appears then to be a relevant issue in all these domains.

Beyond, the modification of the linear and nonlinear optical response of nanocomposite materials by thermal effects represents also an interesting issue to address. Indeed, due to the local electric field enhancement associated with the surface plasmon resonance, gold nanoparticle assemblies have been extensively studied for their optical Kerr effect [13]. In these studies, the influence of thermal phenomena has often been invoked. However, such assessments are most of the time limited to qualitative considerations. Our aim here is not only to provide to the reader the basic theoretical tools for describing linear and nonlinear optical properties of gold nanoparticle assemblies, but also to present original approaches dealing with the interplay between optical and thermal properties which are developed in our group. The main results will be exemplified throughout the paper. Two important practical points will be addressed: The influence of the generalized thermal lens on the third-order nonlinear response, and the hot electron contribution to the gold particle intrinsic third-order susceptibility, including its spectral dispersion and intensity-dependence. The consequences of nanoscale heat transport on the ultrafast relaxation dynamics in nanocomposite media will also be highlighted.

# 2. Linear and nonlinear optical properties of gold nanoparticles: Some elements

For describing the propagation of an electromagnetic wave in an isotropic and homogeneous medium in the linear regime the usual complex optical index $\tilde{n} = n + i\kappa$ will be considered. $n$ is the refractive index and $\kappa$ the extinction coefficient, proportional to the absorption coefficient $\alpha = 4\pi\kappa/\lambda$ ($\lambda$ is the wavelength in vacuum). $\tilde{n}$ is linked with the dielectric function $\varepsilon = \varepsilon_1 + i\varepsilon_2$ through $\varepsilon = \tilde{n}^2$. In this section, we will give some basic elements regarding the optical response of embedded gold nanoparticles which will be useful for the following.

## *2.1 Dielectric function of gold*

The dielectric function of the noble metals has been widely investigated in the past [14–20]. It may include, in the near UV-visible spectral range, the influence of both the quasi-free conduction electrons (*sp* band) and the bound *d* ones. Hence, the total dielectric function of noble metals can be written as the sum of the intraband transition contribution (transitions within the conduction band) and the interband one (transitions from *d* to *sp* bands):

$$\varepsilon_m = \chi^f + \varepsilon^{ib}. \tag{1}$$

### 2.1.1 Intraband contribution: Drude model

The quasi-free electron susceptibility, $\chi^f$, is given by the Drude model [21]:

$$\chi^f(\omega) = -\frac{\omega_p^2}{\omega(\omega+i\Gamma)}. \tag{2}$$

$\omega_p$ is the volume plasma circular frequency and $\Gamma$ is a phenomenological constant accounting for all the collision processes experienced by the conduction electrons, which are considered as independent. The corresponding contributions to $\Gamma$ are then additive (Matthiessen rule). The dominant contribution is the electron-phonon collision one for electron temperatures up to several thousand Kelvin. In the case of a metal nanoparticle, an additional contribution has to be considered due to electron scattering at the particle surface, which is inversely proportional to particle size [22].

### 2.1.2 Interband contribution: Rosei's model

Due to Pauli principle interband transitions present an energy threshold corresponding to the excitation of electrons from the top of the *d* band to states just above the Fermi level ($E_F$) in the conduction band. Thus, below this threshold, $\varepsilon_2^{ib} = 0$. For gold, this threshold lies in the visible (at ~2.4 eV) [17], which explains its color in the bulk state. Contrarily to the intraband contribution to the dielectric function, the interband one cannot be evaluated through a classical approach. Indeed, it is necessary to sum over all possible transitions from an occupied state in the *d* band to an unoccupied state in the *sp* band. The detailed band structure has then to be modeled [14,23–26]. This point has been treated in the 1970s by Rosei and coworkers [24–27].

*Key points for the calculation*

Of course, the interband transition contribution to the dielectric function of gold can be simply calculated by subtracting the Drude contribution from the experimental data of $\varepsilon_m$ following Eq. (1). We describe here the procedure which is usually employed to get the modification of interband transition. One determines the imaginary part of $\varepsilon^{ib}$ first, the real one being subsequently deduced by Kramers-Kronig analysis. In the framework of Lindhard's theory, $\varepsilon_2^{ib}$ writes [21,25]:

$$\varepsilon_2^{ib}(\omega) = \frac{K}{\omega^2} \sum_{i,j} \int_{BZ} |M_{i\to j}(\mathbf{k})|^2 \left[f_i(\mathbf{k}) - f_j(\mathbf{k})\right] \delta\left[\omega_{ij}(\mathbf{k}) - \omega\right] d\mathbf{k}, \tag{3}$$

where $K$ is a constant factor, $\hbar\omega_{ij} = E_j - E_i$ is the energy gap between initial $i$ and final $j$ states, $\mathbf{k}$ is the electron wave vector, $M_{i\to j}(\mathbf{k})$ is the matrix element between states $i$ and $j$, and $f_n(\mathbf{k})$ is the occupation factor of state $n$. The photon wave vector being negligible, the transition is quasi-vertical which imposes energy conservation accounted for by the Dirac function in Eq. (3). Integration is performed over the Brillouin zone (BZ). In gold, the points $X$ (onset at ~1.9 eV [17,27]) and especially $L$ (onset at ~2.4 eV [27]) of the BZ provide the main contributions to *d→sp* interband transitions in the visible. Around these two points the $M_{d\to sp}(\mathbf{k})$ can be considered as locally constant, since the wave number range matching the energy conservation condition is very narrow [24,27]. Thus, for a transition around $X$ or $L$, $\varepsilon_2^{ib}$ is proportional to the joint density of states (i.e. associated with a transition of given energy) which is linked with the densities of states in the *d* and *sp* bands and accounts for the energy and wave vector conservation [24]:

$$J_{d\to sp}^{X,L}(\omega) = \int_{X,L} \frac{2}{(2\pi)^3} \left[f_d(\mathbf{k}) - f_{sp}(\mathbf{k})\right] \delta\left[\omega_{d-sp}(\mathbf{k}) - \omega\right] d\mathbf{k}. \tag{4}$$

Eq. (3) can then be transformed into:

$$\varepsilon_2^{ib}(\omega) = \frac{K}{\omega^2} \left( |M_{d\to sp}^X|^2 J_{d\to sp}^X(\omega) + |M_{d\to sp}^L|^2 J_{d\to sp}^L(\omega) \right). \tag{5}$$

*d*-band states being fully occupied, Eq. (4) can be simplified since $f_d = 1$ whatever $\mathbf{k}$. The conduction band density of states, $f_{sp}$, can be determined by solving Boltzmann equation as will be shown below (§ 3.1.1). The conduction band being isotropic \ref{Ashcroft}, we replace After replacing the integration over wave vector by the equivalent integration over energy. One finally calculates $J_{d \to sp}^{X,L}(\omega)$ which depends on the band structure characteristics, as energies $E_d$ and $E_{sp}$ and effective mass $m_{eff}$, around the BZ points involved. Let us underline that when one aims at determining the modification of $\varepsilon^{ib}$ in gold induced by the absorption of an ultrafast laser pulse, the only significant contribution is that of transitions near point $L$.

*Band structure model*

The band structure of gold has been studied at the beginning of the 1970s [28]. The *d* band group lies a few eV below $E_F$. The *sp* band crosses $E_F$ in the vicinity of the points $X$, $L$ and $\Sigma$ of the BZ. For a given direction of the reciprocal space, it can locally be described by a parabolic branch scheme (quasi-free electron behavior) around the points contributing to the optical response, $X$ and $L$, the dispersion law being then given as $E(\mathbf{k}) = \hbar^2 \mathbf{k}^2 / 2m_{eff}$ [27].

## 2.2  Linear optical response of a single gold nanoparticle

When reducing the volume available for conduction electrons in a metal particle until its size becomes much smaller than the light wavelength, the well-known surface plasmon resonance (SPR) phenomenon arises. Indeed, electrons experience the same homogeneous electromagnetic field and respond collectively (from a quantum point of view, it corresponds to the coherent excitation of electronic transitions within the conduction band). This remarkable property also appears in the expression of the extinction cross section of a metal sphere given by the Mie theory at the dipolar electric order [29]. At resonance the amplitude of the local electric field in the particle, $\mathbf{E}_\ell$, is enhanced as compared to the one of the applied field, $\mathbf{E}_0$. In other words, the modulus of the complex local field factor $f_\ell = E_\ell / E_0$ can be greater than one at the SPR. Numerous recent technological developments in various fields like nonlinear optics [30], sensors, or biomedical imaging are based on this property. In a very simple approach consisting in solving Laplace equation for an isolated metal sphere in an infinite medium one gets for $f_\ell$:

$$f_\ell = \frac{3\varepsilon_d}{\varepsilon_m + 2\varepsilon_d}, \tag{6}$$

where $\varepsilon_d$ is the dielectric function of the surrounding medium. The above equation reveals that $|f_\ell|$ exhibits a resonant behavior as $|\varepsilon_m + 2\varepsilon_d|$ reaches a minimum value. As the existence of such a resonant behavior can be explained by simple electromagnetic considerations using macroscopic parameters as dielectric functions, the origin of the local field enhancement is often known as *dielectric confinement*.

In gold and copper nanoparticles the SPR is located in the visible and presents a lower oscillator strength than in silver, for which the SPR peaks in the near UV. This stems from the significant coupling, in the former case, between bound and quasi-free electron transitions. In addition, the larger the host medium refractive index, the higher the SPR oscillator strength and the lower its frequency. Details regarding the SPR characteristics in noble metal nanoparticles and nanocomposite media can be found in several books [22,31,32]. Let us just point out that the quantum finite size effects (*electronic confinement*) have a large influence on the gold nanoparticle optical response and have been the subject of many theoretical and experimental investigations [22–35].

## 2.3 The case of nanoparticle assemblies

### 2.3.1 Different theoretical approaches

The approach presented above is well suited for an isolated particle or on a pinch for diluted nanocomposite media. In a medium consisting of an assembly of metal nanoparticles spread in a dielectric host the "ensemble effects" have to be taken into account as soon as the density of nanoparticles reaches a sufficient value. Metal volume fraction, $p$, is then the relevant parameter along with the spatial arrangement of the particles. With increasing $p$ mean field effects appear. Further rising $p$ then renders electromagnetic interactions between neighboring particles significant. A large number of effective medium theories have been developed in order to account for such effects. Each one is adapted to a specific material morphology and a specific concentration range. They generally provide an analytical description of the whole medium effective optical response through its effective dielectric function, given the different individual constituent ones. The pioneering and most known one is the Maxwell-Garnett theory (MGT) [36], which supposes no interaction between metal spheres an is thus limited to weakly concentrated media, typically $p < \sim 13\%$ at the SPR (the precise $p$ range for its validity actually depends on wavelength).

Effective medium theories allow to model the electromagnetic response of an inhomogeneous medium while keeping the formalism proper to homogeneous media. However, the hypotheses they impose regarding the shape and size of the particles, their distance, spatial arrangement, concentration and environment in the medium, restrict their validity or lead to develop more and more complex models. An alternative approach can be considered thanks to high power numerical computing possibilities. Given a virtual sample with finite volume, one solves numerically the equations describing mater-radiation interaction in this volume, as Maxwell equations or generalized Mie theory, in order to determine the local electromagnetic field topography. The virtual sample may be chosen as to statistically represent the real global medium. The spatial mean value and dispersion of the field can then be extracted from the results.

We have used such a method to calculate the local field topography in assemblies of spherical gold nanoparticles randomly spread in a transparent host medium. Particle size is supposed to be much lower than the wavelength. This method is based on a multiple-scattering model initially developed for the case of ensembles of scattering spheres, in the formalism of the recursive transfer matrix (*T*-matrix) [37]. The virtual sample is exposed to an external plane wave in the steady regime with linear polarization. The field in a particle is then linked to the applied field and the sum of all fields scattered by the other particles through the associated *T* matrix [38]. The field scattering by each sphere is treated using Mie theory where the maximum multipolar order of the development, $n_{max}$, is imposed. The stability and convergence of the calculation are ensured by the choice of this maximum order, which is as high as the particles are close together and of large size. The dielectric functions of the gold spheres and host medium are assimilated to the ones of the bulk state. The results of these calculations are of course linked with the optical response; for instance, the absorption coefficient of the whole medium is proportional to the mean square of the local field modulus. We will also see further that the latter is involved at the fourth power in the effective third-order susceptibility of nanocomposite materials.

### 2.3.2 Influence of gold concentration

The MGT predicts a red-shift of the SPR with increasing $p$ [22]. This is the consequence of the mean field effect (the MGT does not take into account interactions between particles). One also observes an increase of the SPR absorption band magnitude simply due to the increase of metal amount, the dielectric medium being transparent. We have recently analyzed through simple cases the role of particle interactions on the local field properties depending on the particle arrangement and applied field polarization. For this, the multiple-scattering approach described above has been used. To summarize, the alignment of particles along the applied field direction results in the strengthening and red-shift of the resonance in each particle, whereas an alignment perpendicular to $\mathbf{E_0}$ leads to the decrease of the SPR oscillator strength and a slight blue-shift. These changes increase with reducing inter-particle distance. Their mutual influence acts on both the amplitude and phase of $\mathbf{E_\ell}$. The results obtained in these simple cases, the details of which will be published elsewhere, allow us to analyze more complex situations, describing real samples. We have thus applied our method to the electromagnetic response of a silica host where about 50 to 60 spherical gold nanoparticles, 2 nm radius, are randomly spread. For this, we have generated distributions at a given $p$ value, ensuring that the minimum inter-particle distance is compatible with the convergence of the calculation, the value of $n_{max} = 3$ being imposed as to keep a reasonable computing time. Figure 1 presents the calculated data for two distributions, one (a) having a weak concentration ($p$ =1%), the other (b) a high one ($p$ =22%). The virtual samples are also shown at the same scale in the respective inserts. The quantity reported is the spectral variation of the relative local field intensity ($\sim |f_\ell(\lambda)|^2$) averaged in each particle of the distribution, along with its mean value in the sample. For evaluating this mean value the particles located at the edge of the sample (corresponding to about 30% of the total sample volume) have been disregarded, since their local environment does not match the one the particles in a real sample would have.

One can observe on Fig. 1 the much larger dispersion of the spectra for the larger $p$ value. The presence, in the latter case, of a few very intense resonance bands, shifted to longer wavelengths, is due which actually correspond to a group of very close particles in the virtual sample. Such a behavior has already been highlighted in bidimensional semi-continuous gold films (existence of *hot spots* for the local field) [39–41]. The red-shift of the SPR maximum with increasing $p$ is retrieved for the mean spectrum. One can also notice that in such random distributions there seem to be no noticeable amplification of the mean field intensity; this could stem from the homogeneous distribution of the inter-particle axis orientation relative to the applied field polarization, which should be confirmed by further investigations. It is worth noticing, however, that the medium third-order optical nonlinear response, which is sensitive to the product $f_\ell^2 |f_\ell|^2$ as will be seen in the next section, may exhibit a volume mean value larger than in the case of an isolated particle. One can finally notice on Fig. 1 that the reducing of the inter-particle distance associated with the increase of $p$ leads to the asymmetric broadening of the mean SPR band, absorption being enhanced in its red tail. This fact stems from the increased occurrence of dense particle packing as the one already discussed in Fig. 1(b), and is then fully ascribable to interactions between neighboring particles.

## *2.4    Third-order nonlinear optical response: The optical Kerr effect*

We just give here some elements regarding the third-order nonlinear optical properties of nanocomposite materials, so as to present the issue of our investigations. A more complete analysis of both theoretical approaches and experimental findings has been published in Ref. [30]. The local electric field enhancement in the vicinity of the SPR, beyond its interest in the linear optical regime as described above, makes noble metal nanoparticles attractive for their nonlinear optical response. Indeed, as the latter varies as an integer power of the electric field experienced by mater, the enhancement of $E_\ell$ results in the amplification of the metal nonlinear response as compared to the bulk phase one. The $n^{th}$-order nonlinearity is characterized by the $n^{th}$-order susceptibility, which links the $n^{th}$-order polarization to the $n^{th}$ power of the field [42,43]. In centrosymmetric media the even-order susceptibilities from electric dipole origin vanish. The first non-zero nonlinear susceptibility is then the third-order one, $\chi^{(3)} = \chi_1^{(3)} + i\chi_2^{(3)}$. Disregarding the third-harmonic generation process, the third-order polarization then writes:

$$P^{(3)}(\omega) = 3\varepsilon_0 \chi^{(3)}(\omega) |E(\omega)|^2 E(\omega), \tag{7}$$

where $\varepsilon_0$ is the permittivity of vacuum and $E(\omega)$ is the electric field complex amplitude at circular frequency $\omega$. The simplified notation $\chi^{(3)}(\omega)$ for the susceptibility usually denotes the 4$^{th}$-rank tensor $\chi^{(3)}_{xxxx}(-\omega;\omega,\omega,-\omega)$ if the field is polarized along the $x$ axis. Eq. (7) corresponds to a phenomenon known as the optical Kerr effect, by analogy with the magneto-optic or electro-optic Kerr effects. One can show that it amounts at first order to write the complex optical index as the sum of the linear one (subscript "0") and a term proportional to the wave intensity, $I$, as:

$$\begin{cases} n = n_0 + \gamma I, \\ \alpha = \alpha_0 + \beta I. \end{cases} \tag{8}$$

$\gamma$ and $\beta$ are the nonlinear refraction and absorption coefficients, respectively. They are linked to the complex third-order nonlinear susceptibility through [44–46]:

$$\begin{cases} \chi^{(3)}_1 (\text{esu}) = \dfrac{cn_0}{240\pi^2} \left( 2n_0\gamma - \dfrac{\alpha_0\beta}{2k^2} \right), \\ \chi^{(3)}_2 (\text{esu}) = \dfrac{cn_0}{240\pi^2} \left( \dfrac{n_0\beta}{k} + \dfrac{\alpha_0\gamma}{k} \right). \end{cases} \tag{9}$$

where $k$ denotes the modulus of the wave vector and $c$ the light celerity in vacuum. The components of $\chi^{(3)}$ in Eq. (9) are expressed in the usual electrostatic units (esu), linked with the SI units by:

$$\chi^{(3)}(\text{esu}) = 10^{-8} (c^2/4\pi) \chi^{(3)}(\text{SI}). \tag{10}$$

Let us now examine the optical Kerr effect in nanocomposite materials. As said before, the local electromagnetic field enhancement at the SPR results in the large amplification of the metal nonlinear response which can then be several orders of magnitude higher than the one of the bulk phase [47,48]. This is a direct consequence of the dielectric confinement already evoked in the case of the linear response. Metal nonlinear properties are also affected by electronic confinement: the intrinsic nonlinear susceptibility of a gold nanoparticle is certainly different from that of bulk gold due to finite size effects. We will come back to this point later.

The concept of effective medium can be generalized to the nonlinear optical response of a nanocomposite material [49–51]. Most of the time, such materials contain nanoparticles the nonlinear susceptibility modulus of which is much larger than the dielectric host one. Hence, $|\chi^{(3)}|$ values of the order of 10$^{-8}$ esu and 10$^{-14}$ esu have been reported, respectively, for bulk gold [44] (or gold particles [48]) and different transparent dielectric media in the visible or near-infrared [52–54]. Now assuming that the metal susceptibility, $\chi^{(3)}_m$, is identical for all particles and homogeneous inside each (decoupling approximation) the following relation can be obtained [50,51]:

$$\chi^{(3)}_{eff} = p \left\langle f_\ell^{\,2}(\mathbf{r}) \right\rangle_m \left\langle |f_\ell(\mathbf{r})|^2 \right\rangle_m \chi^{(3)}_m. \tag{11}$$

$f_\ell(\mathbf{r})$ is the local field factor at point $\mathbf{r}$ and the brackets denote the average value in all the metal particles contained in the volume considered. This equation underlines the relevance of the calculation of the local field topography in a nanocomposite medium, as the one presented in § 2.3.2, for analyzing its nonlinear optical response. Equation (11) can further be reduced to a simple analytical expression in the limit case of a dilute medium (low $p$ value) since in this case the local field factor is identical for all particles and is given by Eq. (6). One then obtains [47,49,51,55]:

$$\chi^{(3)}_{eff} = p f_\ell^{\,2} |f_\ell|^2 \chi^{(3)}_m. \tag{12}$$

This expression, together with Eq. (6), is abundantly used in the literature devoted to the optical Kerr effect in nanocomposite media. From Eq. (11) or (12) it is obvious that dielectric confinement induces a local field correction to the nonlinear response of metal nanoparticles, which has several significant consequences [44,56,57]. Since both $f_\ell$ and $\chi_m^{(3)}$ are complex quantities, the medium third-order nonlinear response may present unexpected values. Moreover, the real and imaginary parts of $\chi_{eff}^{(3)}$ may experience strong sign and magnitude variations in the SPR spectral range. Finally, as $|f_\ell|$ exhibits a resonant behavior at the SPR frequency, $|\chi_{eff}^{(3)}| \sim |f_\ell|^4$ is greatly enhanced. This explains both the interest raised by gold nanoparticles for nonlinear optical materials and the high sensitivity of their nonlinear response to every morphological parameter acting on the local field.

After focusing on the effect of dielectric confinement on the third-order nonlinear optical response, let us glance at the intrinsic nanoparticle susceptibility. Surprisingly, very few theoretical works have been devoted to the value of $\chi_m^{(3)}$ in noble metal nanoparticles [57,58] after the pioneering studies of Flytzanis' group who identified and evaluated the different electronic contributions [47,48,56,59]. They first distinguished two pure nonlinear origins corresponding to intraband and interband transitions as in the linear case. The intraband contribution, $\chi_{intra}^{(3)}$, due to a quantum size effect and varying with particle radius $R_p$, is absent in the bulk phase. It is estimated to be negligible as compared with the interband contribution, $\chi_{inter}^{(3)}$. The latter is thought to be mainly imaginary and negative since it corresponds to the saturation of resonant two-level transitions. However, the validity of this argument may be ensured for photon energies larger than the interband transition threshold only. Moreover, the calculations of Hache *et al.* have been carried out in a specific spectral domain, close to this IB threshold in the vicinity of point $X$ of the BZ and to the SPR of gold particles [56]. This actually means that the imaginary character of $\chi_{inter}^{(3)}$ as well as, more generally, the sign and modulus of both electronic contributions, are expected to undergo spectral variations, which is rarely considered in the literature.

As soon as the incident wave intensity is sufficiently high, the modification of the conduction electron distribution induced by photon absorption may result in a significant modification of the optical transition spectrum [60,61]. This is the *Fermi smearing*, which is not a pure electronic nonlinear effect but was demonstrated by Hache and co-workers to amount to an optical Kerr effect contribution [56], that they characterized by the *hot electron* susceptibility, $\chi_{hot\,electrons}^{(3)}$. However, their calculation was restricted to both a specific energy range close to the SPR and a specific excitation pulsewidth (picosecond), a parameter on which the induced electron distribution modification is highly dependent. Thanks to very recent calculations, we will show below that $\chi_{hot\,electrons}^{(3)}$ in fact exhibits strong variations with pulse duration, photon energy and even laser intensity.

It appears then that Flytzanis' group findings should not be taken for granted in being generalized whatever light excitation wavelength, intensity and temporal regime. However, within their restricted applicability domain, they provide a good idea of the nature and order of magnitude of the different contributions to $\chi_m^{(3)}$. Let us summarize their main results: At the SPR frequency of gold nanoparticles the intraband contribution is found to be negligible against the interband one. The latter is mainly imaginary and has a negative sign: $\text{Im}\,\chi_{inter}^{(3)} \sim -1.7 \times 10^{-8}$ esu. For picosecond pulse excitation at the SPR, the hot electron contribution is also mainly imaginary but has a positive sign. Its magnitude is higher than the one of $\chi_{inter}^{(3)}$: $\text{Im}\,\chi_{hot\,electrons}^{(3)} \sim 1.1 \times 10^{-7}$ esu. A compilation and an analysis of several results regarding the value of $\chi_m^{(3)}$, extracted from experimental data obtained on weakly concentrated nanocomposites and reported in the literature, can be found in Ref. [30].

## 3. Dynamics of thermal exchanges in nanocomposite media under pulsed laser excitation

We have been working for few years on the development of models aiming at describing the dynamics of thermal exchanges in nanocomposite media under pulsed laser excitation – depending on pulse duration and energy, particle environment and metal concentration – by accounting for the relevant physical mechanisms involved. Here, we present the main different approaches and the most recent developments, before applying them in the last section to the determination of the role of some thermal effects in the nonlinear optical response of nanocomposites.

The response of a nanocomposite medium to a laser pulse is ruled by a series of different mechanisms, each exhibiting its own dynamics [62–72]: light energy absorption by electrons, redistribution within the conduction electron gas through electron-electron collisions, relaxation toward metal lattice by electron-phonon scattering, and then particle cooling down by heat transfer to the surrounding medium. Note that the short time domain of the relaxation – the first few picoseconds after excitation – has been widely investigated by several groups [62,65,68,70]. We will only give here some basis elements regarding our approaches and disregard any further refinement.

### 3.1.1   Athermal regime: Resolution of the Boltzmann equation

For sake of simplicity, we consider an excitation at photon energy lower than the IB transition threshold. In the opposite case, one should include the effect of Landau damping. When a metal at temperature $T_0$ absorbs an ultrashort light pulse through transitions within the conduction electron gas, the electron distribution is driven out of equilibrium. Internal thermalization ensured by electron-electron collisions leads to a Fermi-Dirac profile at temperature $T_e > T_0$, while electron-phonon scattering converts energy into heat within the particle, the temperature $T_l$ of which rises. Note that the notion of temperature in its strict statistical definition, in a metal object smaller than the heat carrier mean free path, may be questionable. $T_l$ will rather be considered as a measure of thermal internal energy. The *athermal regime* corresponds to the phase during which the electron distribution is off equilibrium. It lasts for a few hundred femtoseconds. It has of course no reason to be considered in a slow excitation regime where electrons and phonons are always at thermodynamic equilibrium.

In the athermal regime, the relevant quantity for describing metal properties is the electron distribution function, $f(E,t)$, where $E$ is electron energy. Its time dependence is ruled by Boltzmann equation which writes:

$$\frac{\partial f(E,t)}{\partial t} = \left.\frac{\partial f(E,t)}{\partial t}\right|_{source} + \left.\frac{\partial f(E,t)}{\partial t}\right|_{e-e} + \left.\frac{\partial f(E,t)}{\partial t}\right|_{e-ph}. \qquad (13)$$

Particles being smaller than the wave penetration depth, the excitation can be considered as homogeneous within the electron gas. Electron diffusion is then ignored in Eq. (13). Moreover, we neglect the influence of the surrounding matrix. This remains justified at short delays after pulse absorption inasmuch as there is no chemical-type interaction at the interface.

In the right part, the first term depicts the evolution of $f$ due to light pulse absorption (source term), while the last two ones denote the variation rate of $f$ due to electron-electron and electron-phonon collisions, respectively. These last years several research groups have developed and carried out more or less refined methods allowing to solve Boltzmann equation in order to describe the athermal regime in noble metals [73–79]. We have elaborated an approach partly similar to the one previously reported in Refs. [77–79]. It is based on a dual-time relaxation approximation for *e–ph* scattering and the Landau theory of Fermi liquids for the *e–e* one.

*Source term*

Dependent on time and electron energy $\varepsilon = E - E_F$, it is proportional to instantaneous power absorbed by metal volume units, $P_{abs}(t)$, as well as to the instantaneous variation of the occupation rate at given $\varepsilon$. This second factor is evaluated by calculating the probability for an electron to be promoted at $\varepsilon$ by absorbing a photon $\hbar\omega$, while subtracting its probability of being promoted at $\varepsilon + \hbar\omega$. The proportionality factor is then determined by imposing the total energy conservation.

*Electron-electron scattering*

This term originates from both the vanishing of electrons of excess energy $\varepsilon$ and the creation of electrons at this energy resulting from screened Coulomb scattering with other electrons. In the weak perturbation case the Landau theory of Fermi liquids allows to apply the time relaxation approximation for describing the first contribution [80]. The corresponding electron lifetime varies as $\tau_{e-e} = \tau_0 E_F^2/\varepsilon^2$: The closer the energy to $E_F$, the weaker the e–e scattering probability, which is a direct consequence of Pauli principle. $\tau_0$ represents the lifetime the electron would have in the absence of this exclusion principle. It is evaluated by considering it as a fitting parameter of the model and comparing with results of relaxation dynamics measurements reported in the literature. Its value then ranges within 0.3–1.0 fs. The second contribution to e–e scattering is calculated from already defined quantities through the expression established by Richie [81].

*Electron-phonon scattering*

The e–ph scattering term in Eq. (13) stems from spontaneous emission, stimulated emission and absorption of phonons. In the last two cases the scattering rate is proportional to the number of available states in the reservoir (phonon population of the reservoir following a Bose-Einstein distribution at $T_l$). The e–ph scattering term is thus split into two contributions, each treated in the frame of the time relaxation approximation. Both depend on the energy transfer rate from the electron gas to the phonon bath, $\dot{q}$. The second one also depends on the number of phonons created until time $t$ and the mean value of the phonon energy (given by the Debye model for the density of states). A nondimensional factor, $S$, accounts for the relative weight of both emission processes and absorption.

*Implementation*

Boltzmann equation is solved numerically by using a finite difference method with a 1 fs time step and a 5 meV energy step. $\tau_0$, $\dot{q}$ and $S$ are free parameters fitted on experimental data from the literature [72,74]. This procedure is rendered possible as these three parameters are independent of $\varepsilon$ and $t$, and as they act on the dynamics in distinct timescales. The typical duration of a calculation of the relaxation dynamics over 10 ps is lower than 5 mn with a personal computer.

*Finite size effects*

The characteristics of the scattering processes driving the evolution of the electron distribution can be modified by finite size effects in nanoscale metal objects. One globally observes the reduction of the e–e and e–ph collision time with decreasing $R_p$ [82–84]. One of the origins of this behaviour lies in the reduction of the screening of the electron-electron and electron-ion Coulomb interactions. The increase of the e–ph scattering rate is also ascribed to the modification of the metal phonon spectrum due to the appearance of vibration modes inherent to confinement [85]. These finite size effects result in a modification of parameters $\tau_0$ and $\dot{q}$.

*Conclusion about the thermal regime*

The resolution of Boltzmann equation allows to account for the athermal regime for the electron distribution and its consequences on the dynamics of the optical response at short time scale after excitation by an ultrashort laser pulse. We will show later an example of such a calculation and its application to the calculation of the hot electron contribution to the third-order nonlinear response of nanocomposite materials. The model used, however, is valid inasmuch as thermal exchanges between particles and their environment remain negligible. When the conduction electron gas has reached internal equilibrium – that is, when it can be described by a Fermi-Dirac distribution at temperature $T_e$ – one enters the *thermal regime*. It is then possible to include the influence of the environment in a three-temperature model.

### 3.1.2 Thermal regime: Three-temperature model

Exchanges between electrons and phonons within metal particles (two-temperature model, 2TM) are very often considered alone to describe the ultrafast optical response dynamics. And yet the presence of the surrounding medium begins to play a significant role just after the first few picoseconds following the excitation, and then cannot be neglected. Moreover, the 2TM does not allow to deal with long-lasting pulse regimes. On the contrary, the three-temperature model (3TM) that we present here enables in principle to carry out the calculation of the dynamics whatever the pulse duration. As the description of this model has already been partly published [45,86], we just recall here the main elements in order to explain the foundation of the application that will be presented in the last section.

*Hypotheses*

For sake of clarity, the conduction electron gas is considered as instantaneously thermalized, i.e. the finite duration of the athermal regime is neglected. This assumption is justified for an ultrashort pulsed excitation of high energy for which the duration of this regime is short, and of course for long pulsed excitation with which the electron gas is always at thermodynamic equilibrium with the metal lattice. It is nevertheless possible to connect the approaches developed for each regime through a modified 3TM, which is out of the scope of this paper.

In the 3TM the thermodynamic characteristics of the different media are likened to the ones of their bulk phase. It would be relatively easy to include in a phenomenological way some effects linked with confinement. The metal/dielectric contact is supposed to be perfect and a possible interface thermal resistance is disregarded. The near field radiative heat transfer is neglected, which is valid in the case of metal nanoparticles the size of which is limited to a few tens nanometers.

*Coupled equations*

Let us consider a nanoparticle with radius $R_p$ isolated in a dielectric host medium. This particle absorbs homogeneously a part of the energy carried by an incident light pulse. The equations to be solved describe the energetic exchanges between the conduction electron gas at temperature $T_e(t)$, metal lattice at temperature $T_l(t)$ and surrounding matrix at temperature $T_m(\mathbf{r},t)$ [86,87]. The input excitation (source term for the electron gas) is described through $P_{abs}(t)$, defined above and depending on pulse characteristics, material absorption and metal concentration. The first equation then depicts the electron energy evolution due to both this source term and the electron-phonon scattering:

$$C_e \frac{\partial T_e}{\partial t} = -G(T_e - T_l) + P_{abs}(t) \tag{14}$$

$C_e = \gamma_e T_e$ is the electron gas specific heat, $\gamma_e$ is a constant depending on metal, $G$ is the *e–ph* coupling constant. The second equation accounts for the evolution of the lattice thermal energy, fed by the *e–ph* coupling and transferred to the surrounding matrix through the interface:

$$C_l \frac{\partial T_l}{\partial t} = G(T_e - T_l) - \frac{H(t)}{V}. \tag{15}$$

$C_l$ is the metal lattice heat capacity, $V$ is the particle volume and $H(t)$ is the instantaneous power released from particle to matrix. Let us now focus on the evaluation of this last term. It has to be treated in a different way depending on the excitation and observation conditions. One has to consider the heat transport mechanisms in the dielectric medium from the interface. In the general case, at small space and time scales, this transport is described by Boltzmann equation or a simplified version; this will be developed later. As soon as the observation timescale (and/or the space scale) is large as compared with phonon lifetime $\tau$ (resp. mean free path $\Lambda$) the use of Fourier's law may nonetheless be considered as reasonable for describing thermal transport. $\tau_{ph}$ and $\Lambda$ are worth, for two usual dielectric media, $\tau_{ph}$~130 fs and $\Lambda$~0.5 nm in silica (SiO$_2$) and $\tau_{ph}$~850 fs and $\Lambda$~5.4 nm in alumina (Al$_2$O$_3$).

*The case of classical diffusive heat transport: Fourier's law*

One adds to Eqs. (14) and (15) the conventional parabolic law of heat diffusion for describing conduction in the dielectric matrix:

$$\frac{\partial T_m}{\partial t} = \frac{\kappa_m}{C_m} \Delta T_m \tag{16}$$

where $C_m$ and $\kappa_m$ are the matrix specific heat and thermal conductivity, respectively. Assuming that $T_l(t) = T_m(R_p,t)$, the function $H(t)$ simply writes:

$$H(t) = S\kappa_m \left.\frac{\partial T_m}{\partial r}\right|_{r=R}, \tag{17}$$

where $S$ is the particle surface.

*Influence of metal concentration*

This model, initially suited for an isolated particle and then for a dilute nanocomposite medium, has been extended to the case of dense assemblies of nanoparticles [86]. Indeed, by preserving spherical symmetry allowing to treat in a relatively easy way the coupled equations and keeping the values of both metal concentration and particle size, we have included the influence of thermal exchanges between neighboring particles in the medium. The use of Fourier's law now imposes the additional condition that the distance between particles be large as compared with $\Lambda$. We have shown that metal concentration plays via these exchanges a crucial role in the temperature dynamics in the case of long pulsewidth excitation, as well as at long pump-probe delays in the case of ultrashort excitation. The main origin of such a dependence lies in the fact that when the heat front emitted by a particle reaches a neighboring one, the temperature gradient at the surface of the latter decreases, which results in the slowing down of its cooling. This is illustrated in the nanosecond excitation regime on Fig. 2 which presents the dynamics of the particle temperature for three different concentrations. One can obviously observe the large increase of the $T_l$ peak value as well as the time shift of this peak with increasing $p$. These results will be used later for evaluating the consequences of particle heating in the optical Kerr response of nanocomposite materials.

*The case of ballistic-diffusive heat transport*

For very short times, small distances and then high metal concentrations, or for matrices with high thermal conductivity, Fourier's law is no longer valid. The Boltzmann transport equation (BTE) then has to be invoked. The question of thermal transport at small time and space scales has been dealt with by several authors [88–94]. An alternative solution to the difficult implementation of the BTE, so called *ballistic-diffusive equations* (BDE) and based on the time relaxation approximation, has been proposed by Chen [90,93]. It is particularly suited for the study of transient thermal phenomena at the nanoscale, for which it has given results quantitatively similar to the BTE ones while requiring much simplified calculations. It consists in splitting in every point of the medium the heat flux $\mathbf{q}(\mathbf{r},t)$ and the internal energy $u(\mathbf{r},t)$ into two distinct parts: $\mathbf{q} = \mathbf{q}_b + \mathbf{q}_d$ and $u = u_b + u_d$. The first one depicts the contribution of ballistic phonons emitted from boundaries and the second one accounts for phonons stemming from scattering processes (or re-emitted after absorption) from other points of the medium.

Chen has mainly adapted this method to the case of a thin film undergoing a sudden temperature rise on one of its sides [91]. In our case, we have included the BDE in the 3TM [95]. For this, the term in Eq. (15) describing the energy exchange at particle-matrix interface is calculated as:

$$H(t) = \int_S \mathbf{q}(\mathbf{r},t)\cdot\mathbf{n}\,ds . \tag{18}$$

Flux integration runs over the whole particle surface, and $\mathbf{n}$ denotes the unit vector normal to the latter, pointing toward the dielectric. By writing the heat flux as the integral of phonon intensity over all energies and all directions of space and developing the diffusive contribution to this intensity as a truncated sum of spherical harmonics to order 1 (method P1, cf. Ref. [94]), and by considering both the energy conservation law and the BTE in the time relaxation approximation for the intensity, Chen has deduced a set of equations (BDE) after averaging conductivity and phonon mean free path over all frequencies [91]. We do not recall his whole reasoning but just write the main equation he obtains:

$$\tau_{ph}\frac{\partial^2 u_d}{\partial t^2}+\frac{\partial u_d}{\partial t}=\nabla\cdot\left(\frac{\kappa_m}{C_m}\nabla u_d\right)-\nabla\cdot\mathbf{q}_b, \tag{19}$$

where $\tau_{ph}$ is the average phonon lifetime. This equation differs from the Cattaneo-Vernotte one (i.e. the hyperbolic diffusion equation obtained by adding an inertial term to Fourier's law as to account for the finite carrier lifetime) from the additional last term relative to ballistic processes. Chen has shown the crucial significance of this term at small space and time scales. He has also highlighted the relevance of using nondimensional parameters for flux, energy, time and space coordinates.

### 3.1.3 Consequences of the nanoscale heat conduction in the host medium

We have implemented the BDE into our 3TM in the case of a spherical metal-dielectric core-shell nanoparticle. This work has been carried out together with S. Volz from the EM2C laboratory, CNRS, Ecole Centrale de Paris, France. A gold core ($R_p$=10 nm) is surrounded by an alumina shell the thickness of which, $d$, is of the same order of magnitude as the phonon mean free path. This particle absorbs (through electron gas excitation) the energy of a 110 fs Gaussian pulse with peak power $P_{abs0}$=1.4×10$^{21}$ W m$^{-3}$. The internal boundary of the dielectric shell (at $r=R_p$) then emits ballistic phonons at particle lattice temperature, $T_l(t)$. The BDE along with appropriate boundary and initial conditions are solved together by a numerical explicit finite difference scheme. Figures 3 and 4 present the result obtained with $d=\Lambda$=5.4 nm in the thermalizing and adiabatic boundary conditions, respectively. The first one corresponds to a full thermalization (temperature $T_0$ imposed at the outer surface) while the second corresponds to a full isolation (cancellation of the heat flux at the outer surface). The quantity reported on these figures is the time evolution of the relative temperature variation of both electron gas and metal lattice. Initial equilibrium temperature is worth $T_0$=300 K. The results provided by the Fourier law have been added for comparison.

In both cases $T_e$ increases very rapidly up to 2200 K at the pulse end. At short times after excitation ($t<5$ ps) energy relaxation is dominated by electron-phonon scattering in metal and remains insensitive to external heat transfer, which explains that electron temperature does not depend on the heat transport mechanism. On the contrary, lattice temperature, which is much lower than $T_e$ due to the much larger specific heat of the phonon bath as compared with the electron gas one, is sensitive to the account for the ballistic regime from the very beginning. After the first picoseconds, electron temperature also exhibits a strong dependence on the heat transport mechanism in the dielectric shell. As can be seen on Fig. 3, in the thermalizing boundary case electrons and phonons are at quasi-equilibrium after about 11 ps. The BDE predicts from then a much slower relaxation than Fourier's law, which is due to phonon rarefaction at metal/dielectric interface. In other words, Fourier's parabolic law overestimates the possible energy redistribution channels due to its inherent assumption of the existence of diffusive processes at any point of the medium and at any time. Moreover, as the input energy remains for a longer time in the particle, the value of $T_l$ predicted by the BDE is higher than the one evaluated with Fourier's law. We have additionally shown that shell thickness has a great influence on the relaxation dynamics [95].

The case of adiabatic boundary conditions (Fig. 4) is quite different since, as the shell thickness is as small as the phonon mean free path, the dielectric medium is very rapidly isothermal due to the homogeneous redistribution of the total input energy within the core-shell nanoparticle. $T_e$ and $T_l$ then reach a plateau after ~10 ps only. A slight difference between the BDE and Fourier's law predictions can nevertheless be still observed, especially for $T_l$ in the first instants after excitation.

In a real experimental situation the relaxation is likely to exhibit an intermediate behavior between these two extreme boundary condition cases, depending on the conductivity of the medium surrounding the core-shell particle. Beyond, the results presented here suggest that thermal transport in the matrix may play a crucial role in the cooling dynamics of ultrafast heated gold nanoparticles and then in the optical response of nanocomposite media.

# 4. Influence of thermal exchanges on the nonlinear optical response

The fast energy input in the electron gas drives the latter out of equilibrium and induces the modification of the transition spectrum implying energy states close to the Fermi level and then the modification of the optical response close to the interband transition threshold. The return of the global medium back to thermodynamic equilibrium via all other processes goes along with the variation of the optical properties due to electron gas cooling down, lattice temperature change, host medium heating and thermal exchange between neighboring particles. Hence, the mechanisms involved in the optical response dynamics mainly originate from thermal effects, at a broad meaning. In this last part, we will describe how temperature variations can be related to the induced variations of the optical response. It amounts in fact to model the mechanisms involved and to evaluate their influence. Depending on the situation (excitation and observation timescales considered, spectral domain, input power,…) it may be possible to focus on a few of them, either the most significant ones, or those the effects of which are to be specifically studied, or treat them in a global manner. We will present here two usual cases, defined relatively to the establishment of the thermodynamic equilibrium between electrons, metal lattice and local environment at the nanoscale. For this, a thin Au:SiO$_2$ film with concentration $p = 8\%$ and thickness $L = 136$ nm will be considered. Particle diameter is fixed at 2.6 nm. In order to partially account for finite size effects on the optical response (broadening and quenching of the SPR absorption band due to electron mean free path limitation [22]) the value of the collision factor $\Gamma$ has been set to 1.1 eV. Let us stress that such a phenomenological way to describe finite size effects in gold nanoparticles would not be sufficient to explain the actual observations for the optical response. Indeed, a realistic interpretation should include the influence of both the free electron spill-out and the skin of reduced polarizability for the core $d$ electrons [35]. The absorption spectrum of our virtual sample, deduced from MGT, is shown on Fig. 5 where the effects of both SPR and interband transitions are clearly identified.

## 4.1 Stationary regime

The stationary thermo-optical regime has been recently studied in our group from both theoretical and experimental points of view [96,97]. Let us here synthesize the key principles and issues in order to apply the findings to one specific aspect of the nonlinear optical response of gold nanoparticles. This regime is rather easy to deal with regarding the thermal aspect since the only knowledge of one temperature is sufficient to describe the thermodynamic state of a particle and its close neighborhood. It is encountered when applying a continuous laser or a pulsed laser the pulsewidth of which is large as compared with the characteristic time of nanoscale heat exchanges. Beyond the issue of overheating by light absorption, this situation corresponds also to thermo-modulation experiments or more simply to cases where the global medium temperature varies macroscopically.

### 4.1.1  Thermo-optical response of gold

The origins of the thermo-optical response of noble metals are multiple and were the subject of numerous investigations in the 1970s, particularly for gold [24,27,98–101]. Various mechanisms lead to the modification of the interband and intraband transitions following a temperature variation, by affecting either directly the electron distribution, or the properties related to the lattice via the increase of the interatomic distance, or the collision processes.

The complex index variation of a material can be linearly linked, to first order, with the temperature variation through the thermo-optical coefficients $d_T n = \partial n/\partial T$ and $d_T \kappa = \partial \kappa/\partial T$ :

$$\Delta \tilde{n}(t) = \left(d_T n + i d_T \kappa\right) \Delta T(t) . \tag{20}$$

The thermo-optical coefficients of gold, averaged in the range 295–670 K, have thus been determined by a precise analysis of the literature, notably by cross-checking different experimental data from thermo-modulation spectroscopy [96].

### 4.1.2  The case of nanocomposite media

The thermo-optical response of a nanocomposite medium depends naturally on that of its constituents. In the manner of the third-order nonlinear optical response of nanocomposite materials (see § 2.4), their thermo-optical response does not amount to the simple volume average of the responses of each of their constituents. Indeed, the local field factor the complex value of which undergoes strong variations around the SPR plays an essential role. As far as the optical properties of an inhomogeneous material can be, under certain hypotheses, described by an effective medium approach like MGT, the latter can be extended to the case of the dependence on temperature. For that purpose, we identify the parameters depending on $T$ (metal and matrix indices, metal concentration) in the expression of the effective susceptibility which we differentiate then. We have thus shown that the effective thermo-optical response in the SPR spectral domain presents a resonant character. The condition of such a resonance appears to be exactly similar to that of the optical response at the SPR. Metal particles present their own thermo-optical response, possibly affected by the interactions with their neighbors. The local field plays then the role of a "complex magnifying glass" by amplifying and warping this intrinsic spectral response in the vicinity of the plasmon resonance. This is particularly perceptible for gold nanospheres where the SPR and the interband transition threshold, at the origin of the main modifications of the optical response, are located at very close photon energies. The consequences are surprising at first sight: The effective thermo-optical coefficients of the global medium are amplified by the effect of local field enhancement, and presents spectral variations in both sign and magnitude very different from those observed for the constituents taken separately. Stationary ellipsometry measurements at different temperatures have confirmed these findings [97]. Additionally, we have been able to infer that the thermal expansion of metal nanoparticles is quenched when embedded in a solid matrix.

### 4.1.3  Application to the nonlinear response: Thermal lens contribution

The principle of thermal lens is as follows. The inhomogeneous intensity distribution in the transverse section of a laser beam induces the heating of the material with a spatial profile correlated to the beam one through optical absorption and energy conversion mechanisms. Due to the material thermo-optical response, the inhomogeneous heating induces an index gradient in the transverse plane. The medium then acts on the beam as a lens, converging or diverging depending on the sign of $d_T n$. Moreover, the beam absorption by the medium may be modulated following the value of $d_T \kappa$. The thermo-optical coefficients then tend to play a role equivalent to that of the nonlinear coefficients for the Kerr effect. This has already been highlighted in various experimental works. By applying the z-scan technique in the time-resolved off-axis scheme Ganeev et al. have demonstrated the presence of a thermal lens effect in the third-order nonlinear response of metal nanoparticles in the nanosecond pulse regime and its absence in the picosecond regime [102]. Liao *et al.* have also invoked thermal effects to explain the discrepancies observed between different pulse temporal regimes [103]. Furthermore, several authors have demonstrated the involvement of thermal lens in the value of $\chi^{(3)}$, stemming from cumulative thermal effects generated at high pulse repetition rates [104–107]. We are thus led to define the thermo-optical contribution to the nonlinear refraction and absorption coefficients, $\gamma_{th}$ and $\beta_{th}$.

We have applied the 3TM together with the results concerning the thermo-optical properties of nanocomposite materials in order to evaluate the contribution of the thermal lens phenomenon to the third-order nonlinear optical response of our virtual Au:SiO$_2$ film in the nanosecond pulse regime. At this timescale the problem is simplified since (*i*) heat diffusion at the macroscale (laser beam waist, about a few tens micrometers) takes several milliseconds to be significant and can thus be neglected during the pulse passage; (*ii*) laser intensity at the beam center is homogeneous within a hundred nanometer scale, which means that all nanoparticles receive the same input energy; (*iii*) at the spatial mesoscale (one particle and its environment) the temperature field can be considered as homogeneous [96]; (*iv*) at the mesoscale the heat conduction characteristic time is worth a few picoseconds, which implies that at the pulse timescale the local medium follows the energy input instantaneously. From the lattice and matrix temperature dynamics calculated in the nanosecond regime by using the 3TM (§ 3.1.2) with typical experimental conditions [30] (pulsewidth: 7.5 ns, absorbed peak power: $P_{abs0} = 5\times10^{17}$ W m$^{-3}$) we have deduced the effective medium temperature increase during the pulse passage, $\Delta\bar{T}$. This parameter is defined as the average temperature increase experienced by the pulse along its passage across the medium, weighted by the pulse intensity profile corresponding to the actual light flux measured in a nonlinear optical experiment [86]:

$$\Delta\bar{T} = \frac{\int_{-\infty}^{+\infty}[T(t)-T_0]P_{abs}(t)\,dt}{\int_{-\infty}^{+\infty}P_{abs}(t)\,dt}. \tag{21}$$

The effective thermo-refractive and thermo-absorptive coefficients, $d_T n$ and $d_T \alpha$, of the Au:SiO$_2$ medium for a given gold concentration are determined as explained in § 4.1.2. The corresponding equivalent thermal nonlinear coefficients are then deduced from:

$$\begin{cases} \gamma_{th} = d_T n\,\Delta\bar{T}/I_{00} \\ \beta_{th} = d_T \alpha\,\Delta\bar{T}/I_{00} \end{cases} \tag{22}$$

where $I_{00}$ is the laser pulse peak energy, calculated from the value of $P_{abs0}$ thanks to medium linear absorption coefficient, air-film transmission, film thickness and metal concentration. In this pulse temporal regime, $\Delta\bar{T}$ is proportional to intensity [86]. Consequently, $\gamma_{th}$ and $\beta_{th}$ do not depend on the $I_{00}$ value. The result of this calculation is presented on Fig. 6 where $\gamma_{th}$ and $\beta_{th}$ are reported as a function of photon energy for the virtual sample described above. Let us describe the main features revealed by these curves. First, the overall order of magnitude of the thermal contribution is fully comparable to that of experimental data obtained in similar conditions. The thermal effect is then likely to compete with the pure nonlinear optical Kerr effect from electronic origin. Secondly, the thermo-optical effect contributes significantly to nonlinear absorption around the SPR, whereas this effect is always disregarded in the literature unlike the refractive part. This lead us to talk rather about a *generalized thermal lens* contribution. Finally, $\gamma_{th}$ and $\beta_{th}$ undergo strong magnitude and sign variations in the plasmon resonance spectral domain. This has interesting consequences as for instance on the thermal lens effect, as illustrated in a schematic manner on Fig. 7: Depending on the incident light wavelength, the same material can behave, either as a converging, or as a diverging lens.

## *4.2 Transient regime*

This regime corresponds in particular to the case where the medium is exposed to ultrashort light pulses, as in a pump-probe experiment or in certain nonlinear optical investigations. Whereas in the stationary regime the mechanisms responsible for the thermo-optical response can be treated in a global manner, they need to be explicitly described in the transient regime as each acts in a different way and follows its own dynamics.

### 4.2.1  Calculation of the optical response dynamics

The optical response of a metal in the visible, near UV or near IR domain, is driven by the properties of quasi-free conduction electrons and bound core electrons as we have seen in § 2.1. The intraband and interband transition spectrum is determined by the knowledge of the conduction electron distribution around the Fermi level, $f(E,t)$, that is the occupation probability of an electron state of energy $E$ at time $t$. $f$ appears then as the relevant parameter to link the metal microscopic behavior to its optical response. When the electron gas is at internal equilibrium (i.e. as soon as $f$ can be given by the Fermi-Dirac statistics) the electron temperature $T_e$ can be employed instead. We have already described above the usual approach to describe the electron distribution dynamics (or that of $T_e$). Once the time evolution of $f$ is determined by this model and the knowledge of several data (as pulse energy and width, wavelength, linear absorption coefficient, initial temperature, particle size, host medium thermal conductivity, metal concentration,…) the Drude (§ 2.1.1) and Rosei (§ 2.1.2) models are used to evaluate the modification of the metal dielectric function. Finally, an electromagnetic calculation method suited to inhomogeneous media – as for instance an effective medium theory or the multiple-scattering approach that we have presented earlier – is carried out to obtain the response of the whole nanocomposite material. One can thus simulate the differential transmission or reflection of a sample in a pump-probe experiment or determine its nonlinear susceptibility.

As an example, the time evolution of the differential transmission $\Delta T/T$ (denoting the relative variation of the sample optical transmittance induced by the pump pulse and measured by the probe one) of the virtual Au:SiO$_2$ film is presented in color levels on Fig. 8 as a function of probe photon energy. The pump pulse duration is 250 fs and its photon energy is 1.55 eV, corresponding to the fundamental frequency of a usual Ti:sapphire femtosecond laser. This result has been obtained by solving Boltzmann equation in the athermal regime as described before (cf. § 3.1.1), including the modification of the *e–e* and *e–ph* characteristic relaxation times due to finite size effects by using the values reported in Ref. [84] for $R_p = 1.3$ nm. One can notice on Fig. 8 the specific signature of gold around the interband transition threshold at ~2.4 eV, corresponding to the change in the conduction electron occupation probability around $E_F$ induced by the pump pulse, amplified and warped by the local field factor at the SPR. As notified before, the influence of the thermal transport mechanism and particularly the importance of the ballistic regime for heat conduction in the matrix is not effective at this short timescale and does not need to be considered in the calculation.

### 4.2.2  Application to the nonlinear response: Hot electron contribution

The hot electron phenomenon is likely to play an important role in the gold nanoparticle intrinsic susceptibility, $\chi_m^{(3)}$, as discussed previously. Let us recall that the only theoretical work devoted to the evaluation of the hot electron susceptibility is the one of Hache *et al.* as discussed in § 2.4. We have used our own approach based on Boltzmann equation or 3TM, depending on pulse duration, to determine the complex value of $\chi_{hot\ electrons}^{(3)}$. From the dynamics of the gold nanoparticle dielectric function provided by the procedure discussed just above it is possible to calculate the effective dielectric function variation during the pulse passage, $\Delta\bar{\varepsilon}(\omega)$, exactly in the same manner as we did in the case of temperature for the thermal lens contribution (cf. § 4.1.3). Indeed, in a conventional nonlinear optical experiment the laser pulse acts as both the pump and probe pulses. $\Delta\bar{\varepsilon}(\omega)$ then writes:

$$\Delta\bar{\varepsilon}(\omega) = \frac{\int_{-\infty}^{+\infty}\left[\varepsilon(\omega,t)-\varepsilon_0(\omega)\right]P_{abs}(t)\,dt}{\int_{-\infty}^{+\infty}P_{abs}(t)\,dt}. \tag{23}$$

The nonlinear optical coefficients are obtained by dividing this value by the peak intensity $I_{00}$, and the nonlinear susceptibility is deduced from Eqs. (9). This procedure has been applied to the virtual Au:SiO$_2$ material previously defined, in the case of 2 ps pulses and with varying incident laser power. Fig. 9 presents with a logarithmic scale the result of the calculations for the modulus of $\chi^{(3)}_{hot\,electrons}$, reported as a function of $I_{00}$ and for selected photon energies. Two important features have to be pointed out on this figure. First, $\left|\chi^{(3)}_{hot\,electrons}\right|$ exhibits very large spectral variations. This stems from two reasons: (*i*) $\Delta\bar{\varepsilon}$ undergoes spectral dispersion at fixed $P_{abs0}$, directly linked with the dispersion of the electron distribution, and (*ii*) $P_{abs0}$ itself strongly varies with $\hbar\omega$ at fixed $I_{00}$, mainly because of the medium linear absorption dispersion (Fig. 5). This second reason – which somehow is of weak physical interest – is nevertheless crucial from a practical point of view: As the energy injected in gold nanoparticles strongly depends on $\alpha_0$ and then on wavelength, the knowledge of the laser beam power (or pulse peak intensity) is far from being sufficient in nonlinear optical measurements on nanocomposite materials. The peak power absorbed by metal volume units, $P_{abs0}$, appears to be a more relevant parameter. The second feature to be underlined on Fig. 9 is that $\left|\chi^{(3)}_{hot\,electrons}\right|$ presents variations with intensity at given photon energy as soon as $I_{00}$ reaches about 10 MW cm$^{-2}$, whereas one should expect a constant value, which is true at low energy only. In the intensity range of usual picosecond lasers (including amplified ones), the hot electron contribution to the gold nanoparticle intrinsic susceptibility is not constant: It is then not a pure third-order nonlinear effect. These findings confirm the care which needs to be taken regarding the generalization of results obtained in the frame of restricted conditions, as already pointed out in § 2.4. Nonetheless, by carrying out similar calculations (not shown here) on another virtual sample free of finite size effect and in a similar temporal pulse regime as to match the model assumptions of Hache *et al.* for the hot electron susceptibility, we obtain a value of $\left|\chi^{(3)}_{hot\,electrons}\right|$ at low intensity and at the SPR frequency close to the one they estimated (1.1×10$^{-7}$ esu) [56]. Moreover, the investigation of the complex value of $\chi^{(3)}_{hot\,electrons}$ reveals that it is purely imaginary and positive at SPR, which was also reported in Ref. [56].

## 5. Conclusion

In spite of the apparent simplicity of the surface plasmon resonance phenomenon, the analysis of the optical response of gold nanoparticle assemblies is far from being straightforward. Indeed, it depends on many parameters regarding both material morphology and excitation characteristics. In this paper, the interplay between the optical and thermal responses of such media have been investigated. On the one hand, we have shown how the interaction of light with matrix-embedded gold nanoparticles can result in the generation of thermal excitations through different energy exchange mechanisms. On the other hand, we have presented how thermal processes can affect the optical response of a nanoparticle assembly. Finally, we have connected both aspects and pointed out their involvement in the nonlinear optical response of nanocomposite media. This has allowed us to take up the challenge of tackling two key issues in the field of third-order nonlinear properties of gold nanoparticles, which had never been carefully addressed until now: The influence of the generalized thermal lens in the long pulse regime and the hot electron contribution to the gold particle intrinsic third-order susceptibility, including its spectral dispersion and intensity-dependence. Additionally, considering the possible inadequacy of conventional macroscopic approaches for describing heat conduction at small time and space scales, we have been able to demonstrate the influence of the heat carrier ballistic regime and phonon rarefaction in the cooling dynamics of an embedded gold nanoparticle subsequent to ultrafast pulsed laser excitation. We have willfully limited our presentation to theoretical approaches, either commonly used, or recently developed in our group, enabling to deal with different situations and including most of the time ensemble effects which can be significant in dense nanoparticle assemblies. Numerous fields need further investigations from the theoretical point of view. Furthermore, the application of these approaches to the analysis of experimental data requires very well defined materials, with a high control level of the morphological parameters, as well as the ability to tune one experimental parameter value only while keeping the others constant. This may represent the adult age of the nonlinear optical investigations on gold nanoparticles.

## Acknowledgements

The financial support of the *Agence Nationale de la Recherche* (program ANR/PNANO 2006, project EThNA) is gratefully acknowledged.

## About the authors

**Bruno Palpant** is associate professor at the Université Pierre et Marie Curie – Paris 6. He conducts his research activities in the Institut des NanoSciences de Paris where he is in charge of the group for nonlinear optics and thermal effects in metal nanoparticles. He got his PhD in 1998 from University of Lyon (France) about quantum size effects in matrix-embedded noble metal clusters, before joining Keio University (Japan) for one year. He has been interested in the linear and nonlinear optical response of noble metal nanoparticles and more recently in their link with thermal conduction at small space and time scales.

**Yannick Guillet** got his PhD in Université Pierre et Marie Curie – Paris 6 in 2007, on the ultrafast dynamics of the nonlinear response of gold nanoparticle assemblies. He now develops femtosecond laser experiments as a post-doctoral researcher.

**Majid Rashidi-Huyeh** is associate professor at University of Sistan and Baluchistan, Zahedan (Iran). He achieved his PhD in 2006 about the influence of thermal effects on the optical response of metal-dielectric nanocomposite materials.

**Dominique Prot** is associate professor of physics at the Université Paris Sorbonne – Paris IV. His research is conducted at the Institut des Nanosciences de Paris (Université Pierre et Marie Curie – Paris 6) on electromagnetic properties of nanocomposite materials and is mainly focused on electric field calculation in matrix-embedded metal nanoparticles.

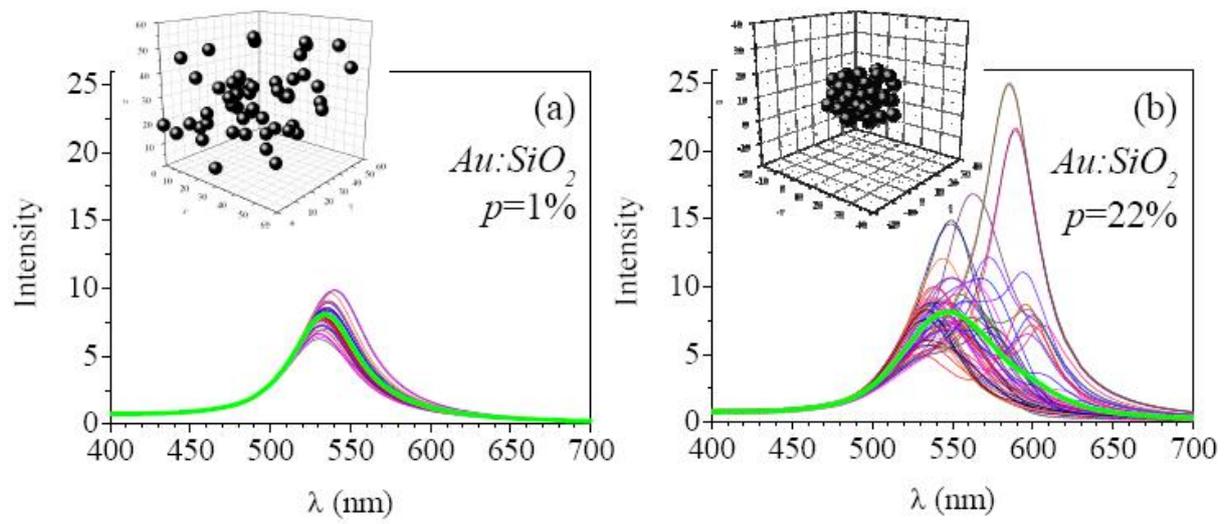

Figure 1. Relative local field intensity averaged in each particle, $|f_\ell(\lambda)|^2$ (thin colored curves), and mean value in the whole sample corrected from edge effects (thick green curve) for a random distribution (insert) of 53 gold nanoparticles (2 nm radius) in a $SiO_2$ matrix, with metal concentration 1% (a) and 22% (b).

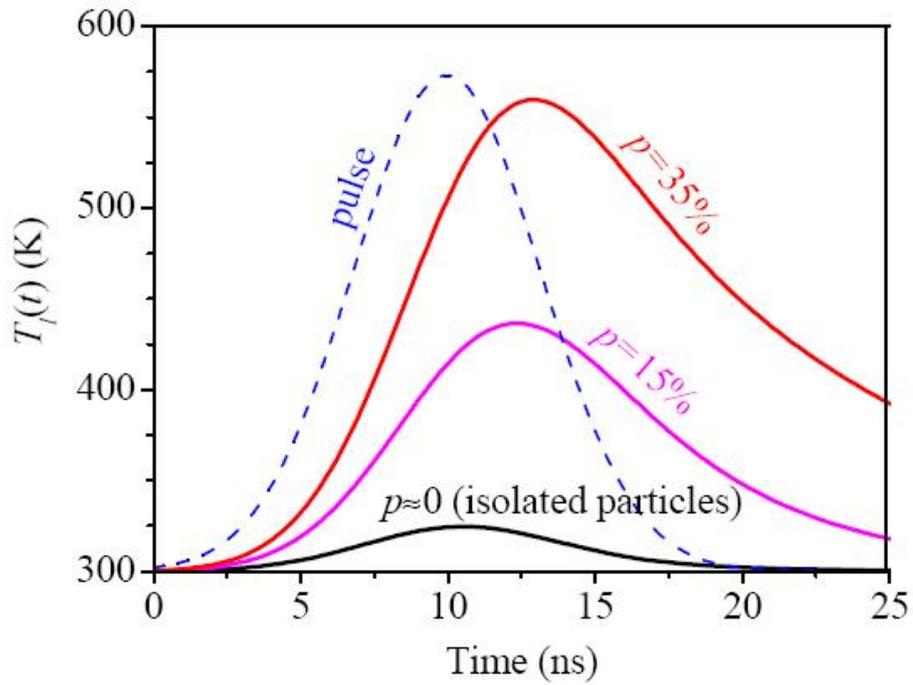

Figure 2. Temperature dynamics of gold nanoparticles ($R_p =1.3$ nm) in silica under pulsed laser excitation (pulsewidth: 7.5 ns, absorbed peak power: $5\times10^{17}$ W m$^{-3}$) for three metal concentrations. The pulse profile is reported in blue (arbitrary units).

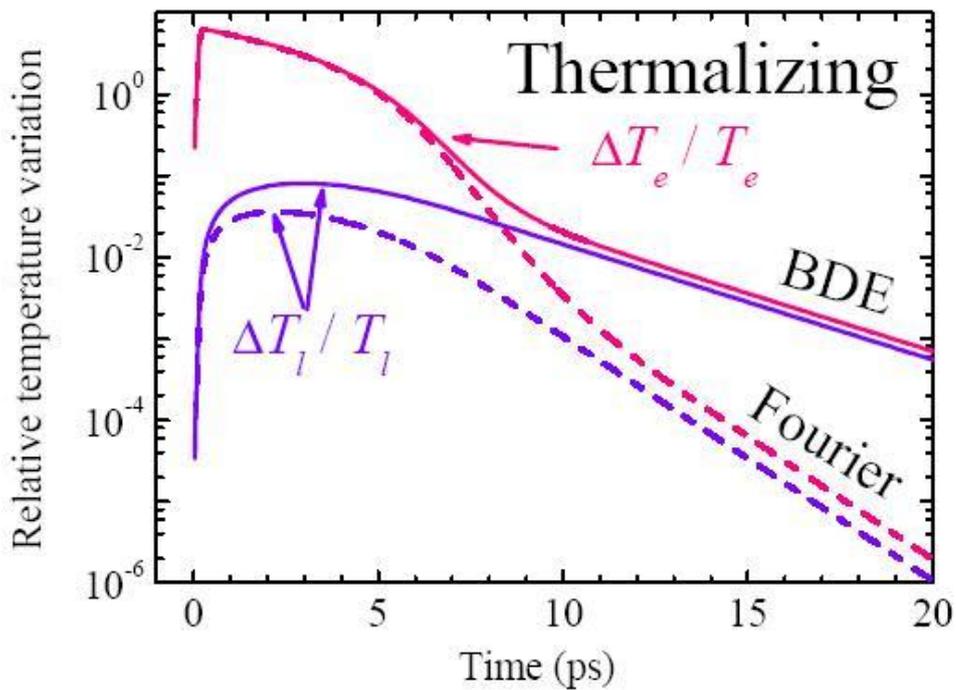

Figure 3: Dynamics of the electron and lattice relative temperature variations induced in a spherical gold nanoparticle (radius: 10 nm) surrounded by an alumina shell (thickness: 5.4 nm) by an ultrashort laser pulse (pulsewidth: 110 fs). Data are obtained using the ballistic-diffusive approximation (solid lines) and Fourier's law (dashed lines). The equilibrium temperature is $T_0 = 300$ K. The boundary condition imposes $T_0$ at the shell outer surface (thermalizing condition). (Collaboration with S. Volz, EM2C laboratory, CNRS, Ecole Centrale de Paris, France.)

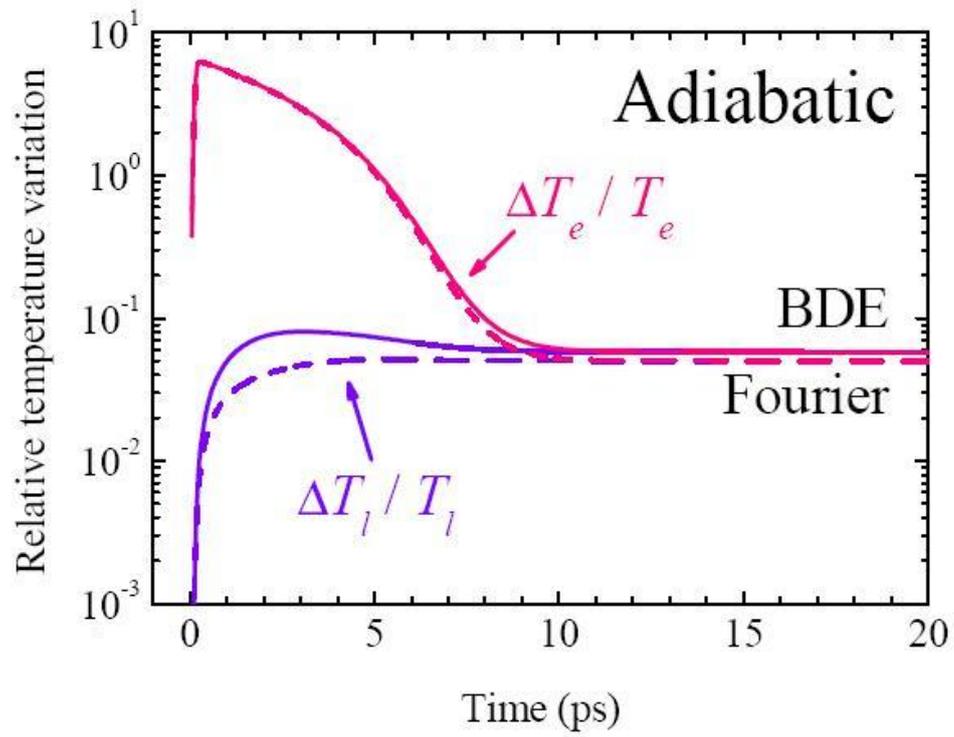

Figure 4: Same as Fig. 3 but the boundary condition consists now in imposing the cancellation of the heat flux at the shell outer surface (adiabatic condition).

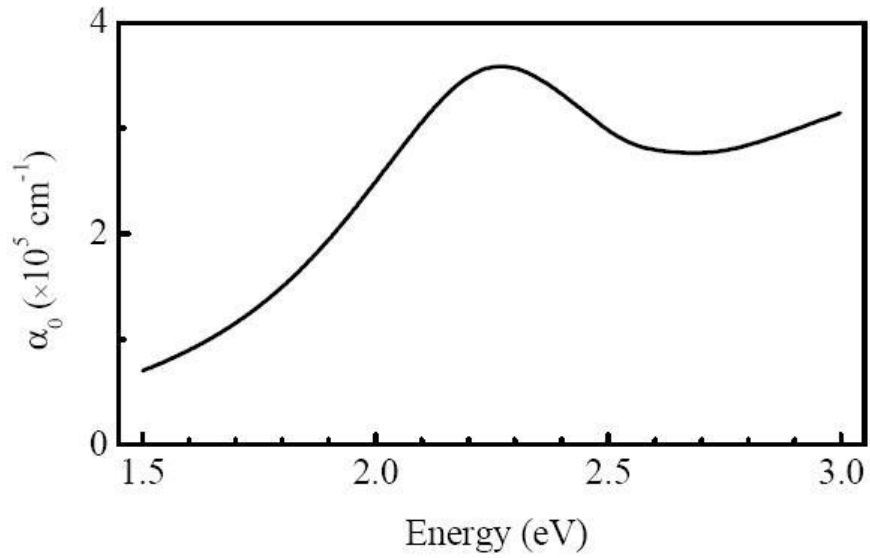

*Figure 5. Linear absorption coefficient of the virtual thin Au:SiO$_2$ film (thickness: 180 nm, p=8%) which will be further considered for the calculation of its nonlinear optical response. The collision factor $\Gamma$ in the Drude model has been set to 1.1 eV to account partially for finite size effects in gold nanoparticles.*

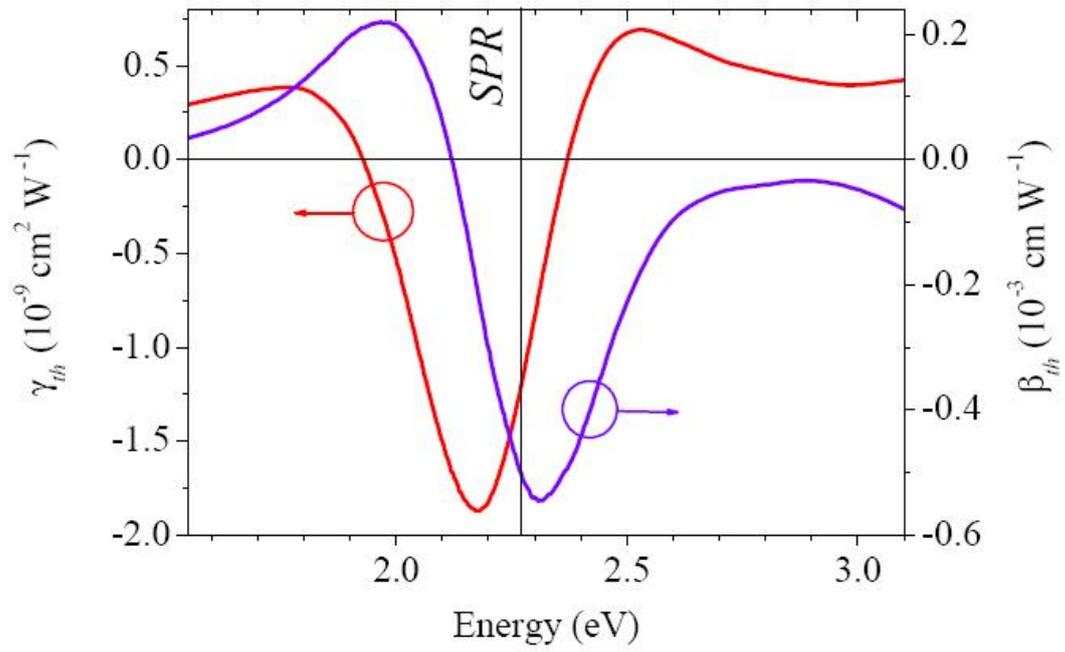

*Figure 6. Calculated spectral dispersion of the thermal contribution to the nonlinear refraction (red) and absorption (violet) coefficients of the virtual thin Au:SiO$_2$ film under 7.5 ns laser pulses.*

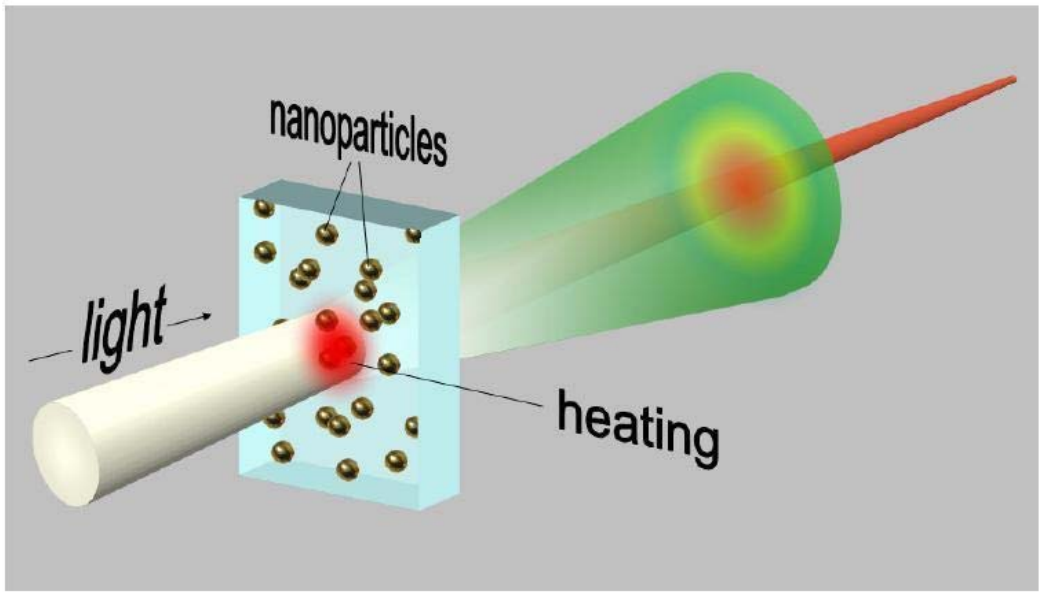

Figure 7. Schematic illustration of the spectral dispersion of the thermal lens phenomenon in a nanocomposite medium.

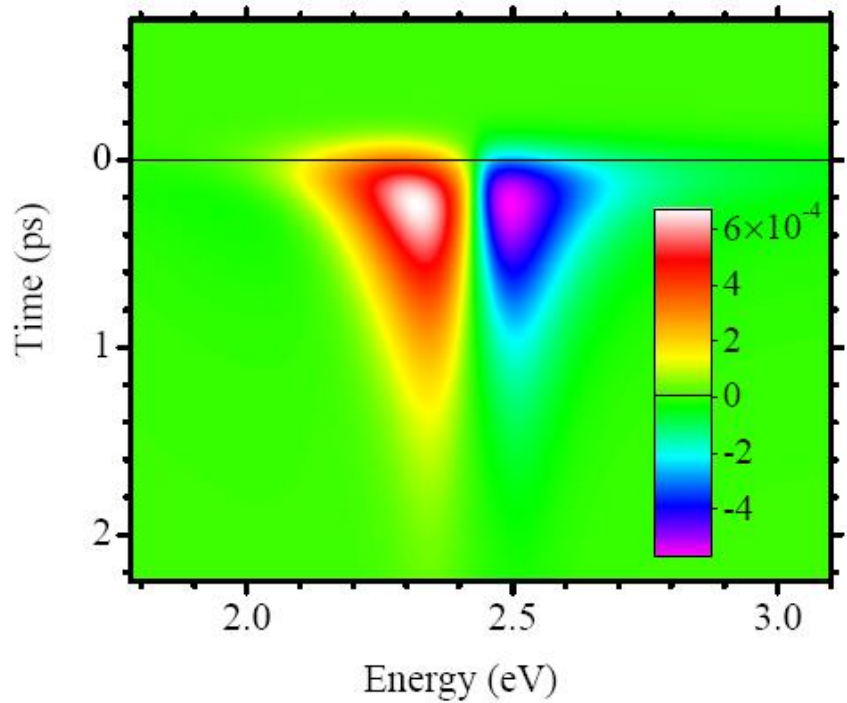

Figure 8. Spectral dispersion of the differential transmission (color level) as a function of pump-probe delay for the virtual thin $Au:SiO_2$ film corresponding to Fig. 5, calculated by solving Boltzmann equation in the athermal regime. Pump photon energy: 1.55 eV; energy absorbed from the laser pulse: $1.4 \times 10^{19}$ W m$^{-3}$; pulse duration: 250 fs. The thin horizontal line at t=0 denotes the instant of pump-probe superimposition.

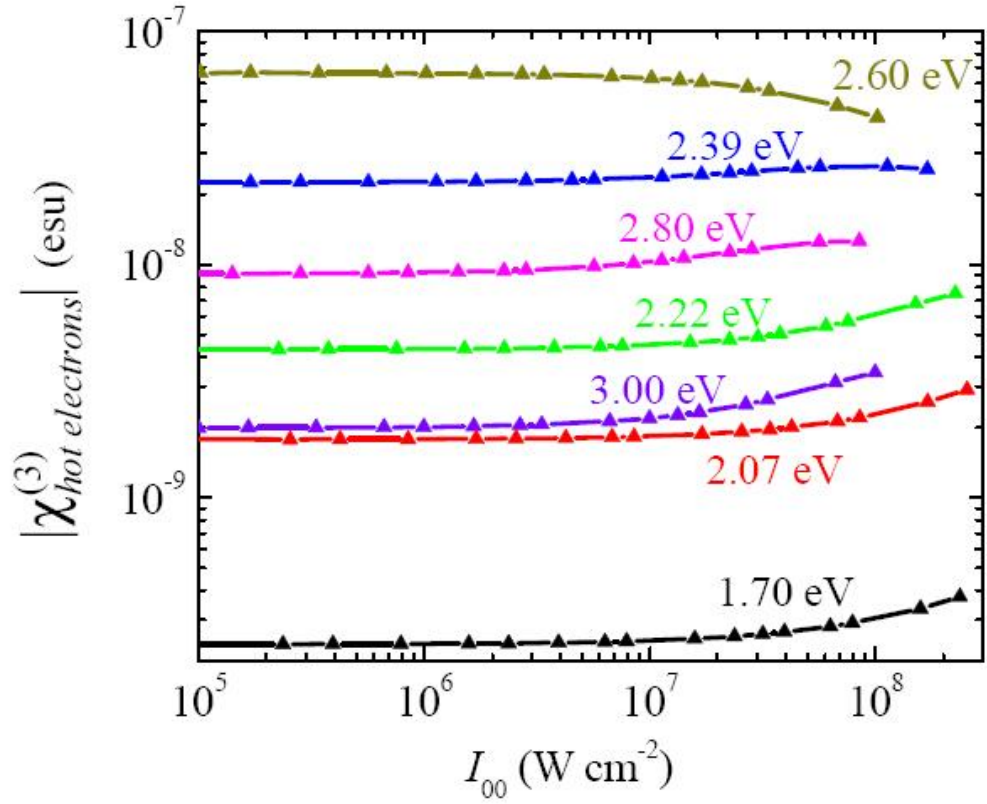

Figure 9. Variation of $|\chi^{(3)}_{hot\ electrons}|$ with incident pulse peak intensity (pulse duration: 2 ps) and at selected photon energies in the virtual Au:SiO$_2$ thin film corresponding to Fig. 5.